# Title: Strong Coupling of a Single Electron in Silicon to a Microwave Photon

**Authors:** X. Mi[1], J. V. Cady[1,†], D. M. Zajac[1], P. W. Deelman[2], J. R. Petta[1*]

**Affiliations:**
[1]Department of Physics, Princeton University, Princeton, NJ 08544, USA
[2]HRL Laboratories LLC, 3011 Malibu Canyon Road, Malibu, CA 90265, USA

*Correspondence to: petta@princeton.edu
†Present Address: Department of Physics, University of California Santa Barbara, Santa Barbara, CA 93106, USA

**Abstract**: Silicon is vital to the computing industry due to the high quality of its native oxide and well-established doping technologies. Isotopic purification has enabled quantum coherence times on the order of seconds, thereby placing silicon at the forefront of efforts to create a solid state quantum processor. We demonstrate strong coupling of a single electron in a silicon double quantum dot to the photonic field of a microwave cavity, as shown by the observation of vacuum Rabi splitting. Strong coupling of a quantum dot electron to a cavity photon would allow for long-range qubit coupling and the long-range entanglement of electrons in semiconductor quantum dots.

**Main Text:** In cavity quantum electrodynamics, light-matter interactions can lead to the coherent hybridization of the quantum degrees of freedom of a photonic cavity and a two-level atom (*1*). A hallmark of cavity quantum electrodynamics is the strong coupling regime, where the coherent coupling rate between the two-level atom and the cavity photon $g_c$ exceeds the photon loss rate $\kappa$ and the atomic decoherence rate $\gamma$. Achieving strong coupling is highly relevant in quantum information science, as cavity photons can be used to mediate long-range qubit interactions (*2, 3*). Strong coupling was first observed in atomic systems using alkali atoms (*4, 5*) and later with optically addressed quantum dots (*6, 7*) and superconducting qubits in the circuit quantum electrodynamics (cQED) architecture (*8*).

Qubits based on electrons confined in semiconductor quantum dots are the focus of intense research efforts due to their scalability and potential for long coherence (*9*). Silicon is a particularly attractive host material for quantum dot qubits, owing to its exceptional spin coherence times (*10, 11*) and highly established fabrication technologies, which in turn have led to the achievement of high-fidelity single-qubit (*12, 13*) and two-qubit (*14*) logic gates. A universal challenge for quantum dot devices is charge noise, which is a dominant dephasing mechanism in both charge and spin qubits (*15, 16*). To date, charge noise has hindered attempts to coherently couple quantum dot devices to microwave frequency photons (*16-18*). Here we report the observation of strong coupling between a single electron in a Si double quantum dot (DQD) and microwave frequency photons in a superconducting cavity, facilitated by highly suppressed charge noise in our device architecture (*19*).

The hybrid silicon-cQED device (Fig. 1A) consists of a half-wavelength ($\lambda/2$) Nb superconducting cavity containing a gate-defined Si DQD. The cavity has a center frequency $f_c$ = 7.684 GHz, loaded quality factor $Q_c$ = 7460, and photon loss rate $\kappa/2\pi$ = 1.0 MHz (Fig. 1B). Three overlapping layers of Al gate electrodes (Fig. 1C) are used to define a DQD in a Si/SiGe heterostructure (Fig. 1D). The electrons trapped in the double-well potential of the DQD are located in an 11 nm thick natural isotopic abundance Si quantum well (QW). The electric dipole interaction couples a single excess electron in the DQD to the electric field of the cavity. We maximize the electric dipole coupling by electrically connecting the plunger gate P2 that is located above the right quantum dot to the superconducting cavity (*16, 20*). Experiments are conducted at a temperature $T$ = 10 mK and sample fabrication details are described elsewhere (*19*).

To demonstrate that the electromagnetic field of the cavity is sensitive to charge dynamics in the DQD, we used a coherent microwave tone with a fixed frequency $f = f_c$ and power $P \approx$ -130 dBm (corresponding to ~3 intracavity photons) to drive the cavity. The amplitude $A$ and phase $\phi$ of the transmitted signal were extracted using a traveling wave parametric amplifier (*21*) and homodyne demodulation technique (*17*). Figure 2A shows the cavity transmission amplitude, $A/A_0$, as a function of the DQD plunger gate voltages $V_{P1}$ and $V_{P2}$. Here $A_0$ is a normalization constant set such that the maximum value of $A/A_0$ is unity with the DQD configured in Coulomb blockade (*17*). A DQD stability diagram is clearly visible in these data, with charge stability islands labeled as ($N_1$, $N_2$) where $N_1$ and $N_2$ is the number of electrons in dots 1 and 2, respectively. At the boundaries of charge stability islands, electron tunneling events damp the electromagnetic field in the cavity, resulting in reduced cavity transmission amplitudes (*17*).

A central figure of merit in cQED systems is the cooperativity $C = g_c^2/\kappa\gamma$, which can physically be interpreted as the ratio of the coherent coupling rate ($g_c$) to the incoherent coupling rates ($\kappa$ and $\gamma$). To estimate this parameter, we focus on an interdot charge transition at which the total number of electrons in the DQD is fixed. Here a single excess electron functions as a charge qubit described by the Hamiltonian $H_a = \frac{1}{2}hf_a\sigma_z$, where $\sigma_z$ is the Pauli matrix (*22*). The qubit transition frequency is $f_a = \sqrt{\varepsilon^2 + 4t_c^2}/h$, where ε is the DQD energy level detuning, $t_c$ is the interdot tunnel coupling and $h$ is Planck's constant (*23*). The cavity is governed by the Hamiltonian $H_c = hf\left(a^\dagger a + \frac{1}{2}\right)$, where $a^\dagger$ and $a$ are the photon creation and annihilation operators, respectively. In addition, the qubit and the cavity are coupled through an interaction Hamiltonian $H_{int} = h(g_c/2\pi)\sin\theta(a^\dagger\sigma^- + a\sigma^+)$ where $g_c$ is the charge-cavity electric dipole coupling rate, $\sin\theta = 2t_c/\sqrt{\varepsilon^2 + 4t_c^2}$, and $\sigma^+$ and $\sigma^-$ are the qubit creation and annihilation operators, respectively (*16, 17*). The total Hamiltonian of the system is the Jaynes-Cummings Hamiltonian $H_{JC} = H_a + H_c + H_{int}$. We performed measurements at the (3,2)↔(2,3) interdot charge transition (Fig. 2B) where the minimum qubit frequency is most closely matched to the cavity frequency $2t_c/h \approx f_c$. Strongly reduced cavity transmission amplitudes $A/A_0 \approx 0$ are observed near the interdot transition (see Fig. 2D). These data give preliminary evidence for highly coherent charge-cavity interactions and a large cooperativity.

To determine the charge-cavity coupling rate $g_c$, we measured $A/A_0$ as a function of ε. Qualitatively, when the qubit-cavity detuning $\Delta/2\pi = f_a - f_c \approx 0$, the eigenenergies $E_{JC}$ of the cQED system (in the subspace where a single quantum of excitation energy is present) are $E_{JC}/h \approx f_c \pm \sin\theta(g_c/2\pi)$ (Fig. 2C). The large detuning between the drive frequency $f = f_c$ and $E_{JC}/h$ results in a strong reduction in the cavity transmission amplitude. Figure 2D shows measurements of $A/A_0$ as a function of ε for three values of $t_c$. With $2t_c/h = 7.72$ GHz, the qubit frequency exceeds the cavity frequency for all values of ε (Fig. 2C) and a single minimum in $A/A_0$ is observed at ε = 0, where Δ is smallest. At lower values of $t_c$, the minimum qubit frequency becomes smaller than the cavity frequency and we observe two minima in $A/A_0$ at values of ε where Δ ≈ 0 (dashed lines in Fig. 2C). Input-output theory is used to account for decoherence effects and to quantitatively fit the measured transmission amplitudes (*17, 18, 24*), with $g_c$ as a free parameter. Inputs to the model are the measured photon loss rate $\kappa/2\pi = 1.0$ MHz and qubit decoherence rate $\gamma/2\pi = 2.6$ MHz (see Fig. 3 for measurements of γ). All three data sets are in excellent agreement with theory, with a best-fit charge-cavity coupling rate $g_c/2\pi = 6.7 \pm 0.2$ MHz (*8*).

Past attempts to reach the strong coupling regime with semiconductor quantum dots have been impeded by background charge fluctuations, which cause temporal fluctuations of the DQD energy levels, resulting in rapid decoherence (*25, 26*). We find that the combination of high quality Si/SiGe heterostructures and a new overlapping gate architecture allow for an appreciable reduction in charge noise (*27, 28*). In these measurements the interdot tunnel coupling is set to $2t_c/h = 7.82$ GHz such that the device is in the dispersive regime with $\Delta \gg g_c$. The cavity phase response Δϕ is measured at $f = f_c$ while gate P1 is driven by an additional microwave tone at frequency $f_s$. The amplitude of the drive is adjusted to minimize power broadening (*27, 28*). The phase response is expected to be $\Delta\phi \approx \pm\tan^{-1}(2g_c^2/\kappa\Delta)$, with positive and negative signs correspond to the qubit being in the ground or excite state, respectively (*27*). When $f_s \approx f_a(\varepsilon)$, the microwave excitation at $f_s$ will increase the excited state population $P_\uparrow$ of the qubit and will lead to more positive values of Δϕ (*27*).

In Fig. 3A, $\Delta\phi$ is plotted as a function of $\varepsilon$ and $f_s$. The frequency dispersion relation of the qubit, corresponding to the resonance condition $f_s \approx f_a(\varepsilon)$, is visible on top of the slowly varying phase response of the cavity. To determine the decoherence rate $\gamma$, we focus on the phase response near $\varepsilon = 0$. Here the qubit is at a "sweet spot" and the energy level separation is first order insensitive to charge noise (*29, 30*). Figure 3B shows $\Delta\phi$ as a function of $f_s$ with $\varepsilon = 0$. The data are fit by a Lorentzian function centered at the qubit frequency $f_a = 7.82$ GHz, in agreement with theory (*27, 28*). The full width at half maximum (FWHM) is $2\gamma/2\pi = 5.2 \pm 0.2$ MHz and is most likely limited by charge dephasing (*27*). This decoherence rate is two to three orders of magnitude lower than previously reported values (*16, 20, 29, 31*) and is pivotal to achieving strong coupling in our system.

In the strong coupling regime, the coherent hybridization of quantum states involving light and matter leads to the emergence of two clearly resolvable normal modes in the cavity transmission spectrum, separated by the vacuum Rabi frequency $2g_c/2\pi$ (*1, 24*). We search for vacuum Rabi splitting with the device tuned to $2t_c/h = 7.68$ GHz, such that the cavity is in resonance with the qubit (at the sweet spot $\varepsilon = 0$). The cavity transmission spectrum $A/A_0$ is plotted as a function of $f$ and $\varepsilon$ in Fig. 4A. At $\varepsilon = 6$ μeV, the transmission spectrum $A/A_0$ is close to that of a bare cavity because of a large qubit-cavity frequency detuning $\Delta$ (red curve, Fig. 4B). However, in the range -2 μeV < $\varepsilon$ < 2 μeV, two maxima emerge in $A/A_0$ at frequencies corresponding to the eigenenergies of the system (overlaid on data). In particular, a pair of distinct peaks is observed with equal heights in the cavity transmission spectrum at $\varepsilon = 0$ μeV, where $f_a = f_c$ (blue curve, Fig. 4B). The peaks are separated by $2g_c/2\pi = 13.4$ MHz, consistent with the value of $g_c/2\pi = 6.7$ MHz extracted from the data in Fig. 2D. The observed normal mode splitting indicates that the strong coupling regime has been reached. This conclusion is supported by independent measurements of $\kappa/2\pi = 1.0$ MHz (Fig. 1B), $g_c/2\pi = 6.7$ MHz (Fig. 2D), and $\gamma/2\pi = 2.6$ MHz (Fig. 3B), yielding a high cooperativity $C = g_c^2/\kappa\gamma = 17$ (*24*). Strong coupling is also achieved in a separate device at the (1,0)↔(0,1) interdot charge transition with [$\kappa$, $\gamma$, $g_c$]/2π = [3.1 MHz, 4.1 MHz, 8.0 MHz]. Our device architecture may be used to achieve spin-cavity coupling through spin-charge hybridization (*18*), which is possible in silicon with the application of an external magnetic field gradient (*32*) or resonant exchange qubits (*33*).

The demonstrated strong coupling between a single electron in Si and a microwave frequency photon paves the way toward the deterministic generation of single photon states (*34*), qubit-photon entanglement (*35*), and long-range coupling of Si qubits (*2, 3*). Due to the rising importance of Si in solid state quantum computing and its technological maturity in the microelectronics industry, the achievement of strong coupling is an important milestone toward the development of silicon-based quantum processors.

**Acknowledgments:** We acknowledge discussions with T. M. Hazard, Y. Liu, S. Putz, M. Reed and J. Stehlik. Research sponsored by ARO grant No. W911NF-15-1-0149, the Gordon and Betty Moore Foundation's EPiQS Initiative through Grant GBMF4535, and the NSF (DMR-1409556 and DMR-1420541). This material is based upon work supported by the Department of Defense under Contract No. H98230-15-C-0453. Devices were fabricated in the Princeton University Quantum Device Nanofabrication Laboratory.

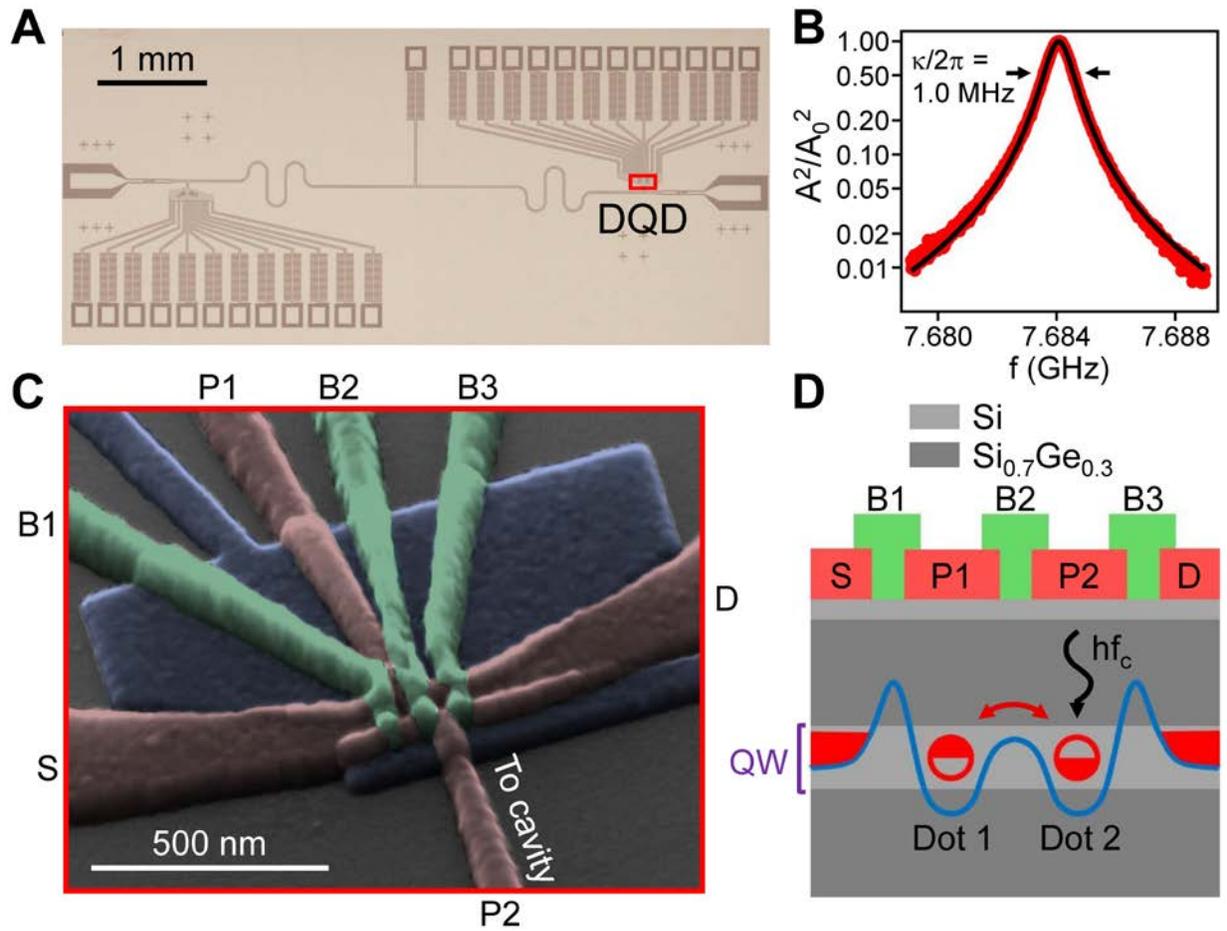

**Fig. 1. Hybrid Si DQD-cQED device.** (**A**) Optical image of a superconducting cavity containing a Si DQD. (**B**) Cavity transmission $A^2/A_0^2$ as a function of frequency $f$ with the DQD in Coulomb blockade and a fit to a Lorentzian with FWHM $\kappa/2\pi = 1.0$ MHz (black line). (**C**) Tilted angle false-colored scanning electron microscope image of the DQD. P1, P2, B1, B2, B3, S, and D are labels for gate electrodes used for confinement of electrons (*19*). (**D**) Schematic cross-section through the DQD gates and Si/SiGe heterostructure. An excess electron is confined in the quantum well (QW) within the double well potential (blue line) created by the gate electrodes. A cavity photon with energy $hf_c$ interacts with the electron.

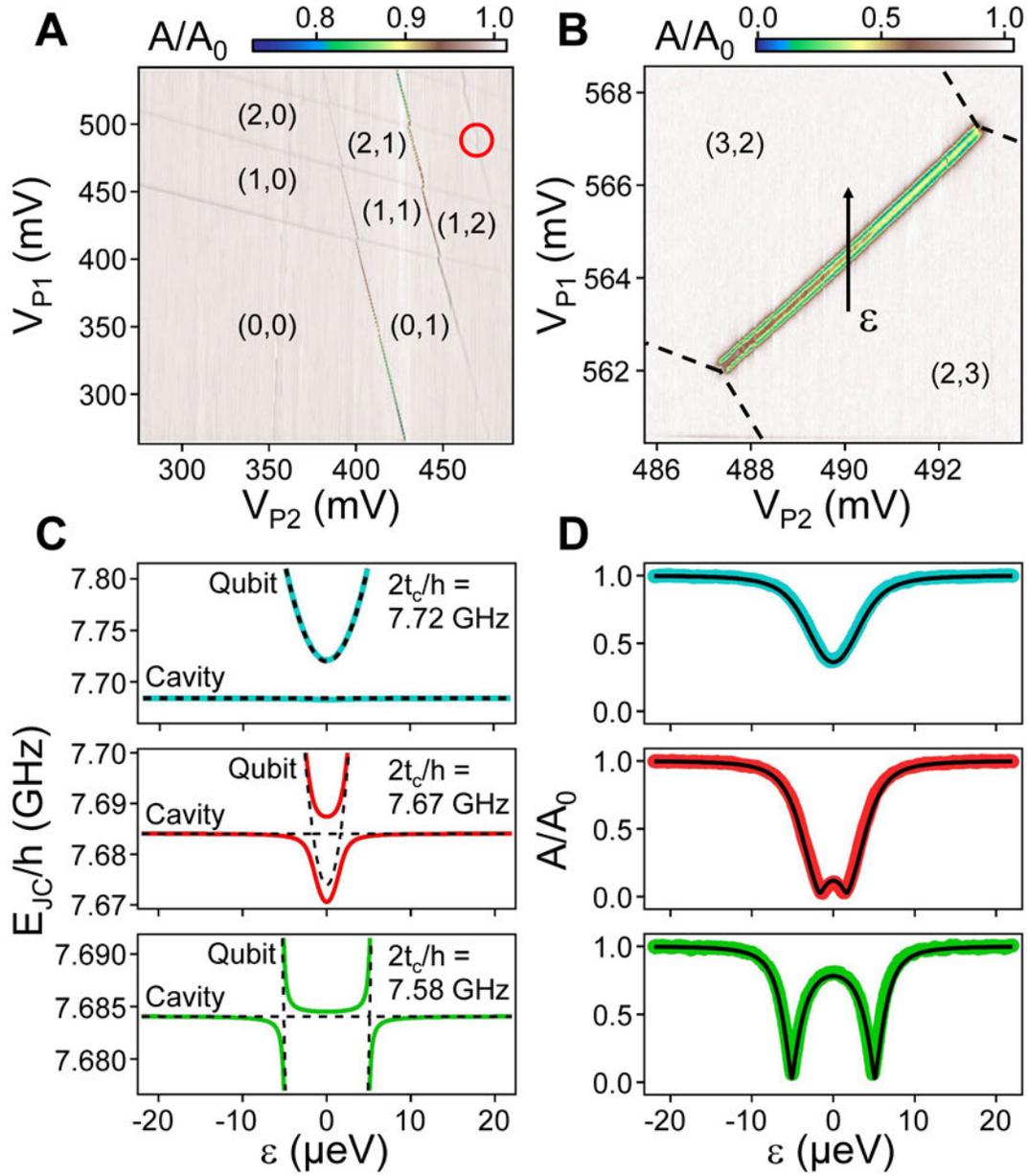

**Fig. 2. Coupling single electrons and photons.** (**A**) DQD charge stability diagram extracted from measurements of $A/A_0$ as a function of $V_{P1}$ and $V_{P2}$, with fixed drive frequency $f = f_c$. (**B**) $A/A_0$ in the vicinity of the (3,2)↔(2,3) interdot charge transition [red circle in (A)], after the DQD is tuned such that $2t_c/h \approx f_c$. Dashed lines mark the boundaries of the stability diagram. (**C**) Eigenenergies $E_{JC}$ of the Jaynes-Cummings Hamiltonian describing the cQED system for three different values of $t_c$, calculated with $g_c/2\pi = 6.7$ MHz (solid lines) and 0 MHz (dashed lines). (**D**) $A/A_0$ as a function of $\varepsilon$ for the values of $t_c$ shown in (C). Black lines are fits to cavity input-output theory.

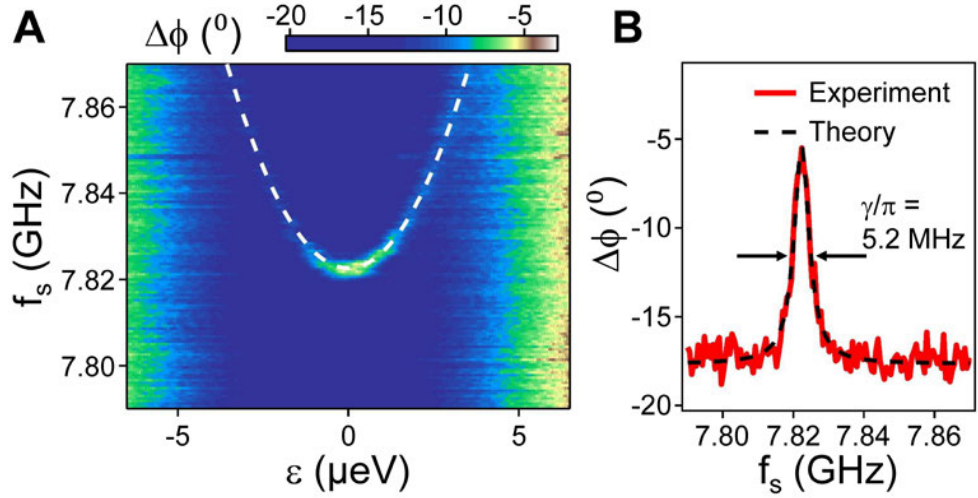

**Fig. 3. Qubit coherence.** (**A**) Cavity phase response $\Delta\phi$ at $f = f_c$ when gate P1 is driven at a variable frequency $f_s$ and power $P_s \approx$ -130 dBm. The qubit dispersion relation, $f_a(\varepsilon)$, with $2t_c/h =$ 7.82 GHz, is overlaid on the data. (**B**) $\Delta\phi$ as a function of $f_s$ at $\varepsilon = 0$ µeV and a fit to a Lorentzian with FWHM $2\gamma/2\pi = 5.2$ MHz (dashed line).

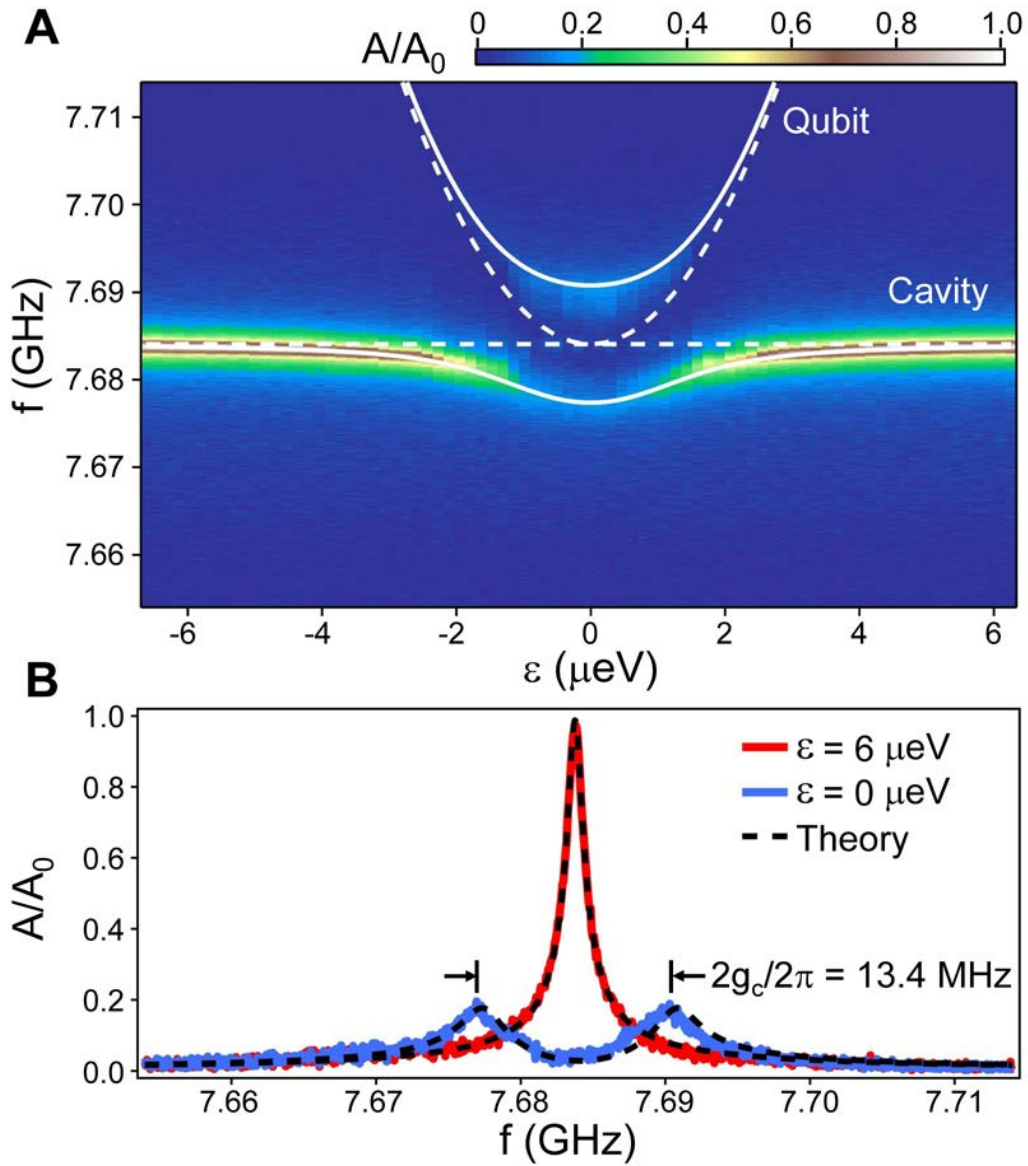

**Fig. 4. Vacuum Rabi splitting.** (**A**) Cavity transmission spectrum $A/A_0$ as a function of $f$ and $\varepsilon$ with $2t_c/h = f_c = 7.68$ GHz. The system eigenenergies are overlaid on the data for the case of no coupling $g_c/2\pi = 0$ (dashed lines) and $g_c/2\pi = 6.7$ MHz (solid lines). (**B**) $A/A_0$ as a function of $f$ at $\varepsilon = 6$ μeV and 0 μeV. Dashed lines are predictions from cavity input-output theory.